\documentclass[acmlarge]{acmart}

\makeatletter
\newcommand{\myconfshort}{\acmConference@shortname}
\newcommand{\myconffull}{\acmConference@name}
\newcommand{\myconfdate}{\acmConference@date}
\newcommand{\myconfloc}{\acmConference@venue}
\AtBeginDocument{
  \fancypagestyle{firstpagestyle}{
    \fancyhead{}%
    \fancyfoot[C]{}%
  }
  \fancyhf{}
  \fancyhead[LO]{\@headfootfont\shorttitle}%
  \fancyhead[RE]{\@headfootfont\@shortauthors}%
  \fancyhead[LE]{\@headfootfont\footnotesize \myconfshort, \myconfdate, \myconfloc}%
  \fancyhead[RO]{\@headfootfont\footnotesize \myconfshort, \myconfdate, \myconfloc}%
  \fancyfoot[C]{}%
}
\makeatother
\acmBooktitle{\conffull\@ (\confshort), \confdate, \confloc}

\usepackage[utf8]{inputenc} 
\usepackage[T1]{fontenc}  
\usepackage{hyperref}    
\usepackage{url}      
\usepackage{booktabs}    
\usepackage{amsfonts}    
\usepackage{nicefrac}    
\usepackage{microtype}   
\usepackage{lipsum}		
\usepackage{graphicx}
\usepackage{natbib}
\setcitestyle{maxcitenames=1}
\usepackage{doi}

\usepackage{comment}
\usepackage{amsmath}
\usepackage{cleveref}

\usepackage{algorithm}
\usepackage{algorithmic}
\usepackage{newfloat}
\usepackage{listings}
\usepackage{textcomp}
\usepackage{xcolor}
\usepackage{mathtools} 
\usepackage{tabularx}
\usepackage{multirow}
\usepackage{multicol}
\usepackage{enumitem}
\usepackage{threeparttable}
\AtBeginDocument{%
 }

\newlist{req}{enumerate}{2}
\setlist[req,1]{label=RQ \arabic*:,ref= \textbf{\arabic*}, leftmargin=*}

\begin{document}

\title[Incentive Alignment for Human-AI Collaboration]{When Thinking Pays Off: Incentive Alignment for Human-AI Collaboration}

\author{Joshua Holstein}
\authornote{These authors contributed equally to this work.}
\email{joshua.holstein@kit.edu}
\affiliation{%
 \institution{Karlsruhe Institute of Technology}
 \country{Germany}
}

\author{Patrick Hemmer}
\authornotemark[1]
\email{patrick.hemmer@partner.kit.edu}
\affiliation{%
 \institution{Karlsruhe Institute of Technology}
 \country{Germany}
}

\author{Gerhard Satzger}
\email{gerhard.satzger@kit.edu}
\affiliation{%
 \institution{Karlsruhe Institute of Technology}
 \country{Germany}
}

\author{Wei Sun}
\email{sunw@us.ibm.com}
\affiliation{%
 \institution{IBM Research}
 \country{USA}
}

\renewcommand{\shortauthors}{Holstein et al.}

\begin{abstract}
Collaboration with artificial intelligence (AI) has improved human decision-making across various domains by leveraging the complementary capabilities of humans and AI. Yet, humans systematically overrely on AI advice, even when their independent judgment would yield superior outcomes, fundamentally undermining the potential of human-AI complementarity. Building on prior work, we identify prevailing incentive structures in human-AI decision-making as a structural driver of this overreliance. To address this misalignment, we propose an alternative incentive mechanism designed to counteract systemic overreliance. We empirically evaluate this approach through a behavioral experiment with 180 participants, finding that the proposed mechanism significantly reduces overreliance. We also show that while appropriately designed incentives can enhance collaboration and decision quality, poorly designed incentives may distort behavior, introduce unintended consequences, and ultimately degrade performance. These findings underscore the importance of aligning incentives with task context and human-AI complementarities, and suggest that effective collaboration requires a shift toward context-sensitive incentive design.
\end{abstract}

\ccsdesc[500]{Human-centered computing~Empirical studies in HCI}
\ccsdesc[500]{Computing methodologies~Artificial intelligence}
\ccsdesc[500]{Human-centered computing~Human computer interaction (HCI)}
\ccsdesc[500]{Human-centered computing~HCI theory, concepts and models}

\keywords{Overreliance, Mechanism Design, AI-Assisted Decision-Making, Appropriate Reliance, Human-AI Collaboration, Human-AI Decision-Making}

\copyrightyear{2026}
\acmYear{2026}
\setcopyright{cc}
\setcctype{by-nc-nd}
\acmConference[FAccT '26]{The 2026 ACM Conference on Fairness, Accountability, and Transparency}{June 25--28, 2026}{Montreal, QC, Canada}
\acmBooktitle{The 2026 ACM Conference on Fairness, Accountability, and Transparency (FAccT '26), June 25--28, 2026, Montreal, QC, Canada}
\acmDOI{10.1145/3805689.3812250}
\acmISBN{979-8-4007-2596-8/2026/06}
\maketitle

\section{Introduction}
Artificial intelligence (AI) is being increasingly introduced to support human decision-making across diverse domains, promising higher decision performance and efficiency. Such performance improvements, however, depend on the complementary capabilities of both partners \citep{Peng_Garg_Kleinberg_2025, Liu_Wei_Liu_Davis_2025}: While AI excels at processing vast amounts of data and identifying complex patterns, humans contribute contextual understanding, ethical reasoning, and the ability to handle novel or ambiguous situations \citep{rastogi2023taxonomy}. However, realizing this complementarity potential in human-AI collaboration fundamentally hinges on the ability of human decision-makers to appropriately rely on AI advice \citep{Schemmer2023Appropriate, guo2024decision}, i.e., knowing when to follow AI advice and when to override it based on their own judgment and expertise.

Yet, achieving this appropriate reliance proves challenging in practice as human decision-makers frequently rely on AI advice, even in situations where their independent judgment would lead to better outcomes \citep{vaccaro2024combinations}, i.e., they ``overrely''. The implications of this overreliance are concerning for high-stakes decisions in domains such as medical diagnosis or criminal justice, where overreliance can have profound consequences for individuals, organizations, and society overall \citep{rudin2019stop}. For example, when decision-makers follow incorrect AI advice, decision quality suffers, and human expertise may gradually erode \citep{lee2025impact}, a pattern supported by evidence that immediate AI collaboration reduces brain activity compared to working independently before collaborating \citep{kosmyna2025your}. To address these challenges, researchers have explored various approaches, yet overreliance remains a persistent problem~\citep{hemmer2021human}.

While existing interventions have shown promise, they primarily focus on individual-level cognitive factors \citep{Riefle2024, spitzer2024don} and information provision like explanations and confidence \citep{buccinca2025contrastive, bansal2021does}, thereby often overlooking the broader environmental factors that shape reliance behavior, as research on human oversight effectiveness highlights that individual-level interventions may be insufficient without addressing the broader socio-technical environment \citep{effectiveness_human_oversight}. This oversight is problematic given that real-world AI-assisted decision-making typically occurs within organizational contexts where performance metrics, rewards, and accountability structures may inadvertently shape reliance behavior. Indeed, meta-analytic evidence from traditional environments shows that monetary incentives significantly enhance task performance by influencing both intrinsic motivation and effort allocation \citep{chen2023cognitive}.
Although the role of incentive designs on human behavior is well-established in organizational behavior research \citep{prendergast1999provision}, their role in human-AI collaboration remains largely unexplored, where the asymmetric cost structure between effortless AI advice without reciprocal obligations and costly independent effort creates novel dynamics. This leads to our research questions that progress from a theoretical analysis to empirical evaluation:

\begin{req}[labelindent=1em, labelwidth=0em, label=\textbf{RQ\arabic*}:, ref=\arabic*]
 \item \textit{What is the role of incentive mechanisms on human reliance in AI-assisted decision making?} \label{rq1}
 \item \textit{How do different incentive mechanisms affect human reliance and performance in AI-assisted decision making?} \label{rq2}
\end{req}

To address these research questions, we draw on an established theoretical framework in human-AI decision-making by \citet{bansal2021most} and identify that systematic overreliance on AI advice cannot be attributed exclusively to the human, but rather stems from an inherent structural cause---a misalignment of incentives embedded in traditional AI-assisted decision-making designs. Even when humans are perfectly rational decision-makers operating with complete information, this structural misalignment systematically biases them toward overreliance, suggesting that the solution lies not in training humans or improving AI explanations, but in modifications to the underlying incentive mechanism. Based on our analysis, we extend this framework by designing an alternative incentive mechanism that introduces monetary rewards for independent decision-making, directly targeting the system-induced overreliance behavior. We then evaluate these incentive mechanisms through a between-subjects behavioral experiment with 180 participants. Our results show that incentive mechanisms significantly reduce overreliance, resulting in higher human-AI team performance. Overall, this work makes three contributions to the field of human-AI collaboration: First, we identify a misalignment of incentives in AI-assisted decision-making as a systemic source for overreliance on AI advice. Second, we propose a new incentive mechanism specifically counteracting the structural misalignment, aiming to reduce human overreliance. Third, we empirically investigate how different instantiations of this incentive mechanism affect human reliance behavior and performance in AI-assisted decision making. We show that dynamic incentives reduce overreliance and improve human-AI team performance, whereas static incentives introduce unintended gaming behaviors that prioritize reward acquisition over performance.

\section{Related Work}
\label{sec:related_work}

Research on human-AI collaboration has extensively investigated the challenge of appropriate reliance, where humans must discern when to follow or override AI advice \citep{Schemmer2023Appropriate}. A substantial body of work has documented the persistent problem of overreliance, where humans systematically accept AI advice even when their independent judgment would yield superior outcomes \citep{schemmer2022meta, hemmer2021human}, particularly problematic in high-stakes decision-making.

To address overreliance, researchers have pursued several intervention strategies. Information-based approaches provide additional contextual information, such as AI confidence scores \citep{bansal2021does, zhang2020effect, uncertainty_expression_user_reliance_trust}, specialized training \citep{spitzer2025human, pinski2023ai}, or explanations of AI reasoning through feature-based \citep{Ribeiro2016, schoeffer2024looks}, example-based \citep{VanderWaa2021, Amitai_Septon_Amir_2024}, or natural language descriptions \citep{buccinca2025contrastive,uncertainty_expression_user_reliance_trust}. Complementary human-centered interventions focus on the development of trust \citep{trust_development} and its calibration \citep{kahr2023seems, Kahr2024} as well as improving mental models of AI capabilities \citep{bansal2019beyond, van_den_bosch_design_2025} or even predicting human behavior \citep{Li_Lu_Yin_2024}, while cognitive interventions employ forcing functions \citep{buccinca2021trust}, confidence alignment techniques \citep{ma2024you}, and prompts encouraging consideration of alternatives \citep{buccinca2025contrastive}.

Although these approaches have demonstrated promising results, they predominantly focus on the AI system or the humans, neglecting the broader socio-technical decision-making environment \citep{effectiveness_human_oversight}, e.g., incentive structures in organizational settings. Incentive structures are a well-established tool in organizations to empower employees toward better outcomes \citep{prendergast1999provision}. Conceptually, it is assumed that monetary incentives can increase effort, which can translate into higher task performance \citep{BONNER2002303}. For example, \citet{STONE1995250} show that performance-contingent incentives induce a longer time to make a decision by examining more information and pursuing decision strategies that ultimately lead to more accurate choices compared to randomly distributed incentives. However, research also highlights the difficulty for incentive mechanisms in capturing all relevant behavioral aspects that can lead to dysfunctional behavioral responses \citep{prendergast1999provision}. In multi-dimensional tasks, employees may optimize for one dimension at the expense of the other if the incentive mechanism only targets a single dimension~\citep{Holmstrom_1991}. 

These dynamics have also been extensively studied in crowdwork settings, where online platforms such as MTurk or Prolific provide a natural laboratory for studying payment structures and their behavioral consequences. A foundational distinction pertains to the type of motivation: \citet{mao2013volunteering} show systematic differences between volunteer and paid crowdworkers in both motivation and contribution quality, whereas \citet{liang2018intrinsic} demonstrate that intrinsic motivation interacts with extrinsic incentives. Specifically, they show that the effect of financial rewards on effort is moderated by workers' intrinsic interest in the task. Regarding the design of payment schemes, evidence suggests that bonuses do not uniformly improve performance. \citet{ho2015incentivizing} establish that performance-based payments raise the quality only for effort-responsive tasks, which aligns with expectancy theory \citep{vroom2015expectancy}, which explains that workers will only exert additional effort if they believe it can be converted into better performance and, in turn, into higher reward. Similarly, \citet{yin2015bonus} show that the decision of whether and how to offer bonuses must be tailored to task characteristics. Beyond effort, incentive structures also influence participation. \citet{patel2023monetary} find that monetary reward structures induce self-selection effects in crowdsourcing contests, altering the composition of contributors and creating trade-offs between participation breadth and contribution appropriateness, resembling an instance of the multi-dimensional incentive problem noted above \citep{Holmstrom_1991}. Finally, financial incentives appear most effective when combined with social mechanisms: \citet{shaw2011designing} find that prompting workers to consider peer responses, alongside monetary payment, yields more accurate judgments than financial incentives alone---a result particularly relevant to the judgment tasks common in AI-assisted decision-making experiments.

Taken together, this body of work suggests that the effectiveness of incentive schemes is highly context-dependent, shaped, for example, by task characteristics, worker motivation, and participation structure. Yet how these dynamics operate when humans decide alongside AI advice remains largely unexplored \citep{ML_and_Mechanism_Design}. Notably, \citet{wu2025human} find that while human-AI collaboration often enhances task performance, it can also simultaneously undermine workers' intrinsic motivation, suggesting that the introduction of AI advice carries incentive-relevant consequences that standard payment schemes do not account for. To the best of our knowledge, only \citet{Wang_Predicting_Human_Behavior} consider the decision-making environment in their work, by varying the rewards associated with each correct decision. However, neither do they study AI advice under the lens of overreliance, nor do they theoretically underpin the decision-making environment; instead, they focus on predicting human behavior. In contrast, we comprehensively analyze and extend an established AI-assisted decision-making framework \citep{bansal2021most} and derive a theoretical model explaining human overreliance on AI advice, which we underline empirically through a behavioral experiment.

\section{Problem Description} 
\label{sec:theory}
In this section, we introduce a formal model of AI-assisted decision-making, which reveals a fundamental misalignment: due to effort costs, humans may rationally defer to AI, leading to overreliance. To address this issue, we propose an incentive-based intervention that introduces a bonus for independent human decisions, designed to realign incentives and mitigate overreliance.

\subsection{Effort-Induced Overreliance}
\label{sec:setting}
We build on the framework of \citet{bansal2021most} to formalize the AI-assisted decision-making setting, where humans receive AI advice but retain control over the final decision. Let \(x \in X \subset \mathbb{R}^n\) denote a task instance and \(Y\) a finite set of possible decisions. For each \(x\), the AI outputs a probability distribution \(h(x)\) over \(Y\), from which the predicted label is \(\hat{y}_{\text{AI}} = \arg\max h(x)\), with associated confidence \(P_{\text{AI}} = \max h(x)\), assuming calibrated scores.

Given this information, the human makes a meta-decision \(m \in \{\text{accept}, \text{solve}\}\): whether to accept the AI's recommendation or solve the task instance independently. If the human solves the task instance, the final decision is \(\hat{y}_{\text{H}}\); otherwise, it is \(\hat{y}_{\text{AI}}\). The environment then returns a utility \(U\) based on the correctness of the final decision.

We model the consequences of decisions using a reward \(\gamma > 0\) for correct outcomes and a penalty \(\beta > 0\) for incorrect ones. In addition, solving a task instance incurs an effort cost \(\lambda > 0\), representing time or cognitive load. The resulting utility outcomes are summarized in \Cref{tab:utility_matrix}.

\begin{table}[h]
 \renewcommand{\arraystretch}{1.3}
 \centering
 \begin{tabular}{c|cc}
  & \textbf{Correct} & \textbf{Incorrect} \\
  \hline
  \textbf{Accept} & $\gamma$ & $-\beta$ \\
  \textbf{Solve} & $\gamma - \lambda$ & $-\beta - \lambda$ \\
 \end{tabular}
 \caption{Utility outcomes in AI-assisted decision-making.}
 \label{tab:utility_matrix}
\end{table}

Let \(P_{\text{H}} = P(\hat{y}_{\text{H}} = y \mid m = \text{solve})\) denote the human's probability of being correct when solving the task instance. Assuming the human is a rational agent who trusts the AI, their meta-decision maximizes expected utility. Thus, the human will accept the AI’s advice if and only if:
\begin{align*}
\mathbb{E}[U(m=\text{Accept})] &\geq \mathbb{E}[U(m=\text{Solve})] \\
P_{\text{AI}} \gamma + (1 - P_{\text{AI}})(-\beta) &\geq P_{\text{H}} (\gamma - \lambda) + (1 - P_{\text{H}})(-\beta - \lambda)
\end{align*}
By rearranging the expected utility comparison, the condition under which a human accepts the AI's advice becomes:
\begin{align*}
 P_{\text{AI}} &\geq P_{\text{H}} - \frac{\lambda}{\gamma + \beta}
\end{align*}
This condition reveals a structural misalignment: the human is incentivized to defer to the AI whenever its probability of being correct exceeds the human's, adjusted downward by the term \(\frac{\lambda}{\gamma + \beta}\). 
Since \(\lambda > 0\), the adjustment is strictly positive, lowering the threshold for accepting AI advice. As a result, the human may rationally choose to follow the AI even when \(P_{\text{H}} > P_{\text{AI}}\). While this misalignment decreases with negligible effort or sufficiently high stakes (i.e., large \(\gamma + \beta\)), it vanishes only when \(\lambda = 0\).

In effect, the human does not defer because the AI is more accurate, but because the \textit{effort to override is not worth the marginal gain}. This demonstrates that overreliance can emerge \textit{even under perfect information}, driven not by irrationality but by the incentive structure itself.

\subsection{Incentive Alignment via Bonus Mechanism}
\label{sec:new_setting}

To address this \textit{misalignment of incentives}, we propose a novel incentive mechanism for human-AI decision-making that aims to \textit{mitigate overreliance} on AI. Specifically, we introduce a new \textit{bonus} term, denoted by \(\theta\), awarded to human decision-makers who invest effort and correctly solve a task instance independently. The updated utility structure with the bonus is summarized in \Cref{tab:utility_matrix_new}.

\begin{table}[h]
 \renewcommand{\arraystretch}{1.3}
 \centering
 \begin{tabular}{c|cc}
  & \textbf{Correct} & \textbf{Incorrect} \\
  \hline
  \textbf{Accept} & $\gamma$ & $-\beta$ \\
  \textbf{Solve} & $\gamma - \lambda + \textcolor{orange}{\theta}$ & $-\beta - \lambda$ \\
 \end{tabular}
 \caption{Updated potential outcomes with the bonus \(\theta\).}
 \label{tab:utility_matrix_new}
\end{table}
\vspace{-1.1em}
Incorporating this bonus $\theta$, the decision rule becomes:
\begin{align*}
\mathbb{E}[U(m=\text{Accept})] &\geq \mathbb{E}[U(m=\text{Solve})] \\
  P_{\text{AI}} \gamma + (1 - P_{\text{AI}}) (-\beta)
  &\geq P_{\text{H}} (\gamma - \lambda + \textcolor{orange}{\theta}) + (1 - P_\text{H}) (-\beta - \lambda) 
\end{align*}

Rearranging terms yields a new threshold condition:
\begin{align*}
  P_{\text{AI}}
  &\geq P_{\text{H}} + 
    \frac{\textcolor{orange}{\theta} P_{\text{H}}-\lambda}{\gamma + \beta}  
\end{align*}

This expression highlights the incentive correction introduced by \(\theta\), i.e., when \(\theta = \lambda / P_\text{H}\), the right-hand term vanishes, eliminating the incentive misalignment. Accordingly, \(\theta\) should scale with the required effort \(\lambda\), providing stronger incentives when solving is more costly or when the human success probability \(P_\text{H}\) is low. By compensating for effort and/or task difficulty, the bonus \(\theta\) encourages independent problem-solving and mitigates overreliance on AI advice, thereby promoting more balanced and effective human-AI collaboration.

The theoretical framework reveals that effective human-AI collaboration requires reconceptualizing incentive structures beyond simple rewards for correct decisions. Rather than providing only a fixed reward $\gamma$, we introduce bonuses $\theta$ that allow for the redistribution of the total available compensation between rewards ($\gamma$) and bonuses ($\theta$) for independent decision-making, while keeping the maximum payouts fixed.

This enables different bonus allocation strategies ranging from \textit{static bonus} across all instances to \textit{dynamic bonus} that varies the magnitude and availability of bonuses across different task instances based on where the expected utility of human effort is largest. In the following, we present a behavioral experiment to empirically investigate the effectiveness of these mechanisms in mitigating overreliance on AI.

\section{Behavioral Experiment Design}
\label{sec:experiment_design}
To evaluate the effect of our proposed incentive mechanism, we conducted a behavioral experiment  with 180 participants performing an image classification task, approved by the university's  institutional review board. We employed a between-subjects design with three conditions: a  \textit{Baseline} condition replicating standard AI-assisted decision-making, a \textit{Static  Bonus} condition providing a fixed bonus for independent correct classifications, and a  \textit{Dynamic Bonus} condition providing a variable bonus targeted at instances where human judgment is most valuable. Below, we describe the experimental design, including the task and dataset, the AI system, and the overall setup of the behavioral study.

\textbf{Dataset and Task.} We selected a task that requires no prior domain knowledge but remains sufficiently challenging to demand meaningful effort. As noted by \citet{fugener2021will}, such generic tasks often yield more generalizable insights, ensuring that observed effects are attributable to the incentive mechanism rather than task-specific confounds, and establishing a foundation for future studies to build upon with increased complexity. To this end, we used a multi-class image classification task based on the ImageNet-16H dataset \citep{steyvers2022bayesian}, which comprises 1,200 images across 16 classes with phase noise distortion applied. This distortion introduces varying levels of difficulty for both human and AI classification. In addition to ground-truth class labels, the dataset includes annotations from six independent human annotators per image. We used annotator disagreement as a proxy for human probability (\(P_\text{H}\)) to correctly solve the instance: higher disagreement indicates greater difficulty in classification, while lower disagreement suggests the task is easier for humans.

\textbf{AI System.} We fine-tuned a DenseNet-161 \citep{huang2017} pre-trained on ImageNet \citep{Russakovsky2015} on the distorted images. Details are left to the Appendix.

\begin{figure}[h!]
 \centering
 \includegraphics[width=0.6\linewidth]{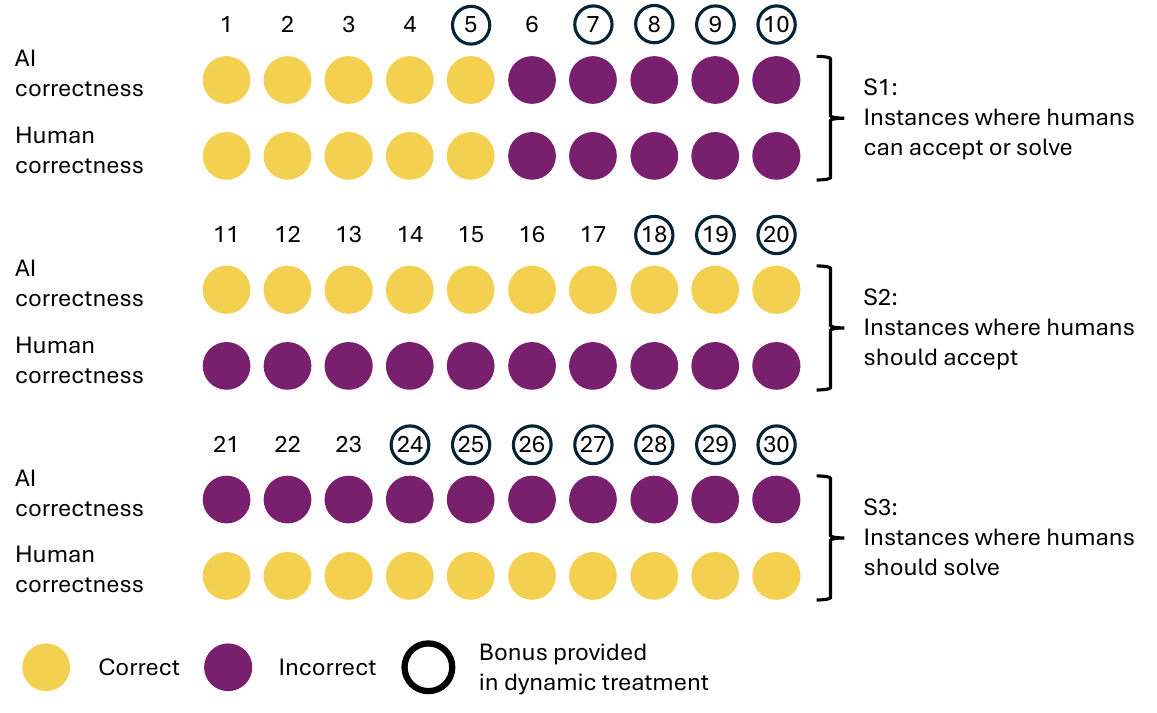}
 \caption{Overview of the 30-instance selection scheme.}
 \label{fig:instance_overview}
\end{figure}

\textbf{Instance Selection.} For the experiment, we aimed to simulate authentic real-world scenarios where the relative strengths of humans and AI vary across task instances. We selected 30 instances representing three complementary scenarios (\Cref{fig:instance_overview}): (S1) both humans and AI perform similarly, (S2) AI outperforms humans, and (S3) humans outperform AI. To determine the relative performance of humans and AI, we used existing human annotator performance from the dataset and compared this with whether the AI prediction was correct or incorrect to strategically sample instances. This yielded 10 instances per scenario (S1-S3), enabling evaluation of how incentive mechanisms perform across different complementarity conditions. Complete selection criteria appear in the Appendix.

\textbf{Treatments.} Based on our theoretical framework, we derived the following allocation of reward (\(\gamma\)) and bonus (\(\theta\)) across treatments (for full derivations see the Appendix): In the \textit{Baseline} condition, participants received a fixed reward for correct classifications (\(\gamma = 0.06\)) and no bonus for solving independently (\(\theta = 0.00\)), replicating the standard incentive structure our framework identifies as misaligned. In the \textit{Static Bonus} condition, the reward was reduced (\(\gamma = 0.03\)) and a fixed bonus (\(\theta = 0.03\)) was introduced for correctly solving any instance independently. The bonus amount was adjusted to equalize expected payouts between always following AI advice and always solving independently. Finally, in the \textit{Dynamic Bonus} condition, participants again received the same reduced reward (\(\gamma = 0.03\)), but the bonus (\(\theta = 0.06\)) was selectively applied only to instances where the AI was not expected to outperform the human, specifically, those with low calibrated AI confidence. No bonus was given when the AI was sufficiently confident, thereby focusing human effort where it is most valuable. In both bonus conditions, the bonus was awarded only when participants chose to solve independently and answered correctly, even if their answer then matched the AI’s advice. All three conditions were calibrated to yield the same maximum payout of \pounds1.80.

To guide effort allocation, AI confidence scores were shown in all conditions, grouped into four bins (very low, low, high, very high). A five-minute time limit was enforced to simulate effort constraints (\(\lambda\)), encouraging strategic decision-making under time pressure, where solving independently consumes time that could otherwise be used to complete additional instances, thereby creating an opportunity cost consistent with our theoretical model and similar to real-world scenarios.

\textbf{Participants.} 
We recruited 180 participants through Prolific, with 60 participants assigned to each treatment condition. Eligibility was restricted to U.S. residents fluent in English. The sample consisted of 43.6\% male participants, with an average age of 39.6 years.
The median completion time was 12 minutes, in which all participants received a performance-independent participation payment of \pounds1. In addition to the performance-independent base payment of \pounds1, participants earned an average of \pounds0.81 in performance-based incentives, including both (\(\gamma\)) and bonus (\(\theta\)) components. This results in a total average compensation of \pounds1.81, which exceeds Prolific's recommended remuneration rate.

\begin{figure}
  \centering
  \includegraphics[width=0.9\linewidth]{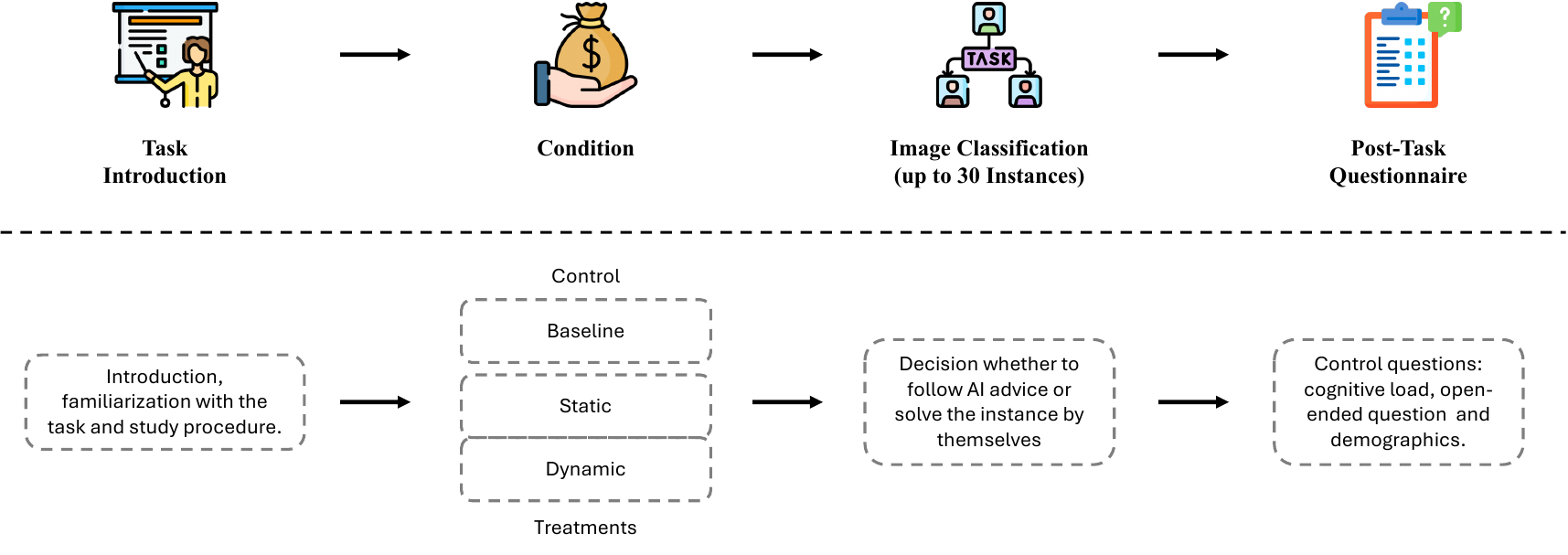}
  \caption{Procedure of the behavioral experiment.}
  \label{fig:procedure}
\end{figure}

\textbf{Study Procedure.} Participants were randomly assigned to one of the three treatment conditions and introduced to the task, including the bonus structure and time limit (\Cref{fig:procedure}). A comprehension check ensured understanding of the incentive mechanisms; participants who failed this twice were excluded from the study. Next, participants completed a guided tutorial explaining the task interface, including AI advice, confidence levels, timer, and bonus eligibility. After two training instances, they proceeded to the main task, consisting of 30 instances to be completed within a five-minute global countdown timer visible throughout the task. On each instance, participants were simultaneously presented with the distorted image, the AI's predicted class label, its confidence level, and any applicable bonus (see \Cref{fig:task_interface} in the Appendix). Participants then chose to either follow the AI's recommendation or solve the task independently, the latter advancing them to a classification screen with a 4$\times$4 grid of all 16 possible classes. The AI's advice was then no longer displayed on the classification screen. Following the task, participants completed a post-task questionnaire measuring cognitive load, along with treatment-specific open-ended questions regarding how AI confidence and the payment structure influenced their decisions. An optional feedback section concluded the study.

\textbf{Metrics.} Our analysis examined three primary measures across experimental conditions: task completion rates (i.e., the number of instances processed within the time limit), human-AI team performance, and reliance behavior. Additionally, we measured cognitive load by using the NASA-TLX scale \citep{hart2006nasa}.

\section{Experimental Results}
\label{sec:results}

\subsection{Task Completion}
\label{sec:task_completion}

\begin{figure}[h!]
\centering
\includegraphics[width=0.75\linewidth]{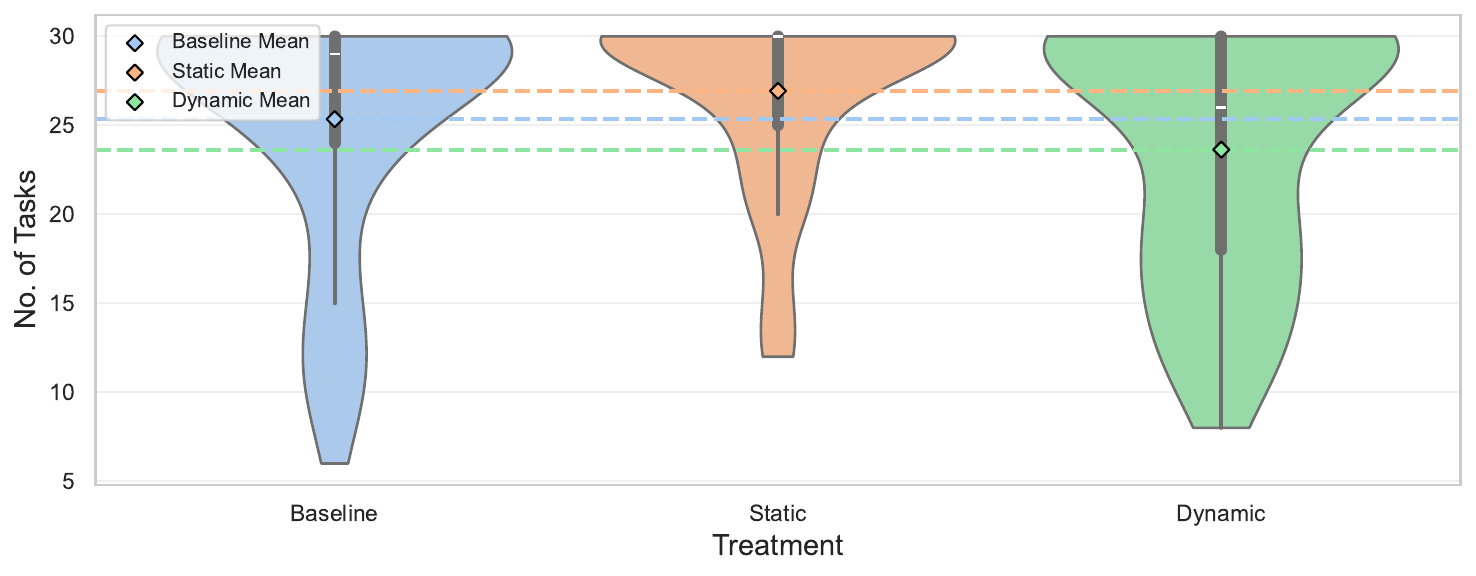}
\caption{Distribution of processed instances across experimental conditions.}
\label{fig:task_completion}
\end{figure}

We begin by examining the effect of the treatments on task completion, measured as the total number of instances classified within the allotted time. Our analysis reveals differences in task completion rates across experimental conditions, with static bonus participants completing the most instances ($\mu_{s} = 26.93$), followed by baseline participants ($\mu_{b} = 25.35$), and dynamic bonus participants completing the fewest instances ($\mu_{d} = 23.63$) (see \Cref{fig:task_completion}). As Shapiro-Wilk tests confirmed non-normal distributions and Levene's test revealed heterogeneous variances across conditions, we employ a Kruskal-Wallis test, which confirms statistically significant differences in completion rates ($H = 7.5$, $p = 0.023$), with static bonus participants processing approximately 14\% more instances than dynamic bonus participants.

\textbf{Mediation Analysis.} To explore the mechanisms underlying reduced task completion rates in the dynamic treatment, we conducted a mediation analysis examining whether cognitive load mediates the relationship between incentive structures and task completion. The analysis reveals that dynamic bonuses significantly increase participants' cognitive load compared to baseline ($\beta = 0.403, p = 0.027$), which in turn reduces task completion ($\beta = -1.094, p = 0.023$), with cognitive load accounting for approximately 26\% of the total effect (see the Appendix for full details).

These systematic differences in completion rates have implications for measurement precision in our subsequent analyses. Participants who completed fewer instances yield less reliable estimates of individual performance and reliance behavior due to smaller sample sizes. Therefore, an unweighted analysis would treat all participant means as equally precise, which is inappropriate given the heteroskedasticity induced by the design. To account for this heterogeneity, we apply inverse variance weighting \citep{hartung2011statistical} in our analyses of human-AI team performance and reliance behavior, assigning greater weight to participants with lower variance and more observations. To further enhance the robustness of these weights, we apply winsorization at the 5th and 95th percentiles to the weight distributions within each treatment group \citep{tukey1977exploratory} as detailed in the Appendix. All subsequent treatment comparisons are conducted using weighted independent samples t-tests comparing each treatment against the baseline condition, with Bonferroni correction applied to control for multiple comparisons. This pairwise approach is preferred over omnibus tests as our interest lies in contrasting each incentive condition with the baseline rather than in differences among the incentive conditions themselves. Therefore, we treat the baseline condition as the reference point against which each incentive mechanism is evaluated.

\subsection{Human-AI Team Performance}
\label{sec:results_accuracy}

\begin{figure}[h!]
 \centering
 \includegraphics[width=0.75\linewidth]{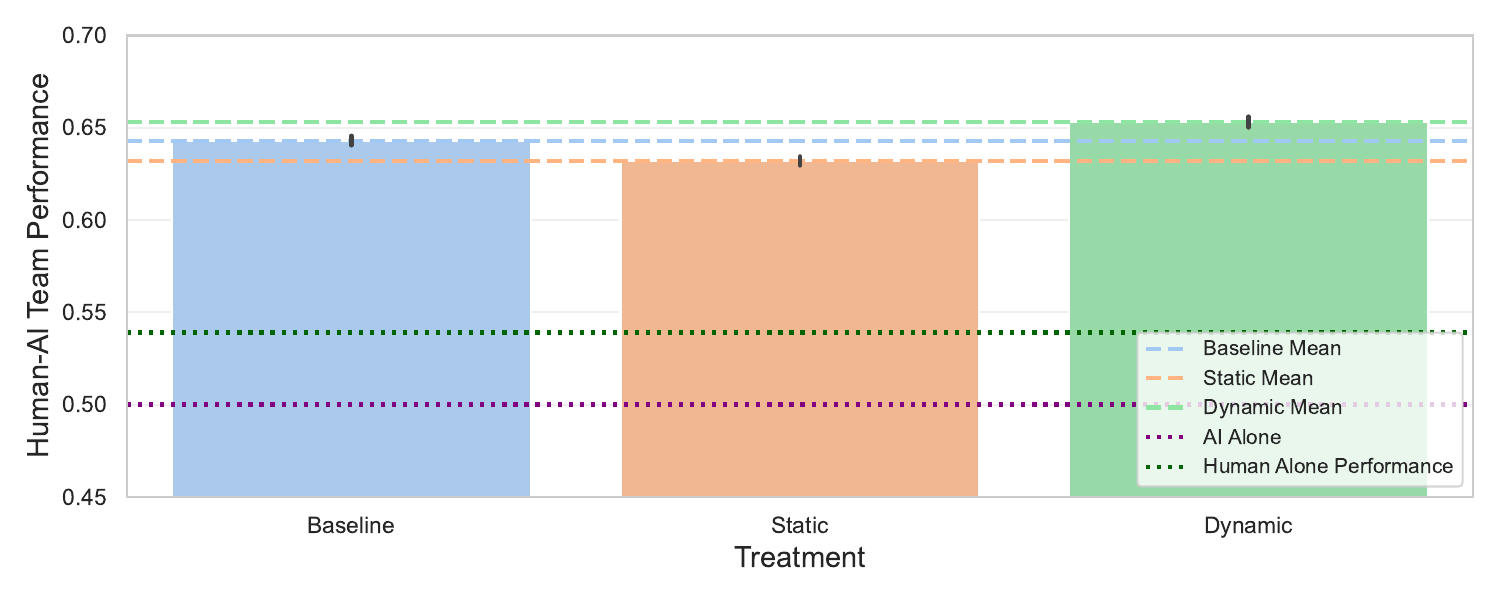}
 \caption{Weighted average human-AI team performance across treatments with 95\% confidence intervals. Reference lines show AI-only performance (purple) and human-only performance (dark green) derived from dataset ground truth labels.}
 \label{fig:hai_performance}
\end{figure}

Next, we assess the impact of incentive mechanisms on overall human-AI team performance. As Shapiro-Wilk tests confirmed normality of the performance distributions across all three conditions, we apply the weighted t-tests introduced above. Our analysis reveals contrasting effects: the static and dynamic bonus treatments lead to opposite shifts in accuracy relative to the baseline. More specifically, participants in the static bonus condition achieved lower accuracy on average than those in the baseline group (\(\mu_{s} = 0.632\), \(\mu_{b} = 0.643\); \(t = -6.01\), \(p < 0.001\), \(d = -0.10\)). In contrast, participants in the dynamic bonus condition exhibited significantly higher accuracy than baseline (\(\mu_{d} = 0.653\); \(t = 5.04\), \(p < 0.001\), \(d = 0.09\)) (see \Cref{fig:hai_performance}). While these aggregate differences are modest, they mask important heterogeneity across instance types, as we detail below.

\textbf{Targeted Accuracy Gains from Bonus Placement.} To further understand how bonus placement drives performance differences, we analyze instance-level accuracy within the dynamic treatment by comparing participant performance to the baseline on the \textit{same} instances. When bonuses are available in the dynamic condition, participants significantly outperform baseline counterparts (\(\mu_{d} = 0.537\), \(\mu_{b} = 0.524\); \(t = 2.83\), \(p = 0.009\), \(d = 0.07\)). In contrast, when no bonus is available, no significant difference is observed (\(\mu_{d} = 0.796\), \(\mu_{b} = 0.786\); \(t = 1.517\), \(p = 0.259\), \(d = 0.031\)).

This pattern suggests that the observed performance gains are not due to a general motivational effect, such as participants trying harder throughout the tasks, but rather stem from the targeted placement of bonuses. If motivation alone were driving the improvements, we would expect accuracy gains even on non-bonus instances. Instead, the effects appear only where bonuses are present, indicating that the mechanism works by directing human effort to where it is most needed.

\begin{table}[h]
\centering
\resizebox{0.6\linewidth}{!}{%
\begin{threeparttable}
\begin{tabular}{lccc}
\toprule
\textbf{Instance Type} & \textbf{Baseline} & \textbf{Static } & \textbf{Dynamic} \\
\midrule
(S1) Human \& AI perform equally & 0.613 & 0.601* & 0.599* \\
(S2) AI outperforms humans & 0.814 & ~~~~0.841*** & ~~~~0.871*** \\
(S3) Humans outperform AI & 0.499 & 0.482* & ~~~~0.573*** \\
\bottomrule
\end{tabular}
\begin{tablenotes}
 \item \textit{Note:} * \textit{p} $<$ .05; ** \textit{p} $<$ .01; *** \textit{p} $<$ .001.
\end{tablenotes}
\end{threeparttable}
}
\vspace{0.5em}
\caption{Team accuracy across complementarity conditions.}
\vspace{-1em} 
\label{tab:complementarity}
\end{table}

\Cref{tab:complementarity} further contextualizes these findings by presenting weighted accuracy across three performance scenarios: (S1) human and AI perform equally well, (S2) AI outperforms humans, and (S3) humans outperform AI. In S1, both treatments slightly but significantly reduce performance (\(d = -0.07\) for both), suggesting unnecessary deliberation. In S2, both improve performance, with dynamic bonuses yielding the greatest gains (\(d = 0.13\) for static and \(d = 0.26\) for dynamic). The most interesting and theoretically important pattern emerges in S3, where static bonuses slightly degrade performance (\(d = -0.06\)), while dynamic bonuses increase accuracy by 7.5 percentage points (\(d = 0.25\))---precisely the instances where human input is most valuable. These instance-level patterns represent the core empirical contribution, in which dynamic incentives successfully redirect human effort toward where it yields the greatest collaborative benefit.

Together, these findings demonstrate that while the static bonus applied uniformly can lead to misallocated effort and reduced overall performance, the dynamic bonus improves accuracy by incentivizing effort where human input is most valuable. By selectively aligning incentives with instance-level difficulty, the dynamic bonus enhances human-AI complementarity. Having established these accuracy effects, we now turn to examine how the different incentive mechanisms influence reliance behavior, specifically, whether bonuses shift participants' willingness to accept or override AI advice.

\subsection{Reliance Behavior}
\label{sec:reliance_behavior}
Finally, we examine participants' reliance behavior to assess whether our incentive interventions effectively reduced overreliance, as suggested by our theoretical framework. We define reliance as the proportion of instances where participants accepted the AI’s suggestion without attempting to solve the problem themselves.

\begin{figure}[h!]
 \centering
 \includegraphics[width=0.75\linewidth]{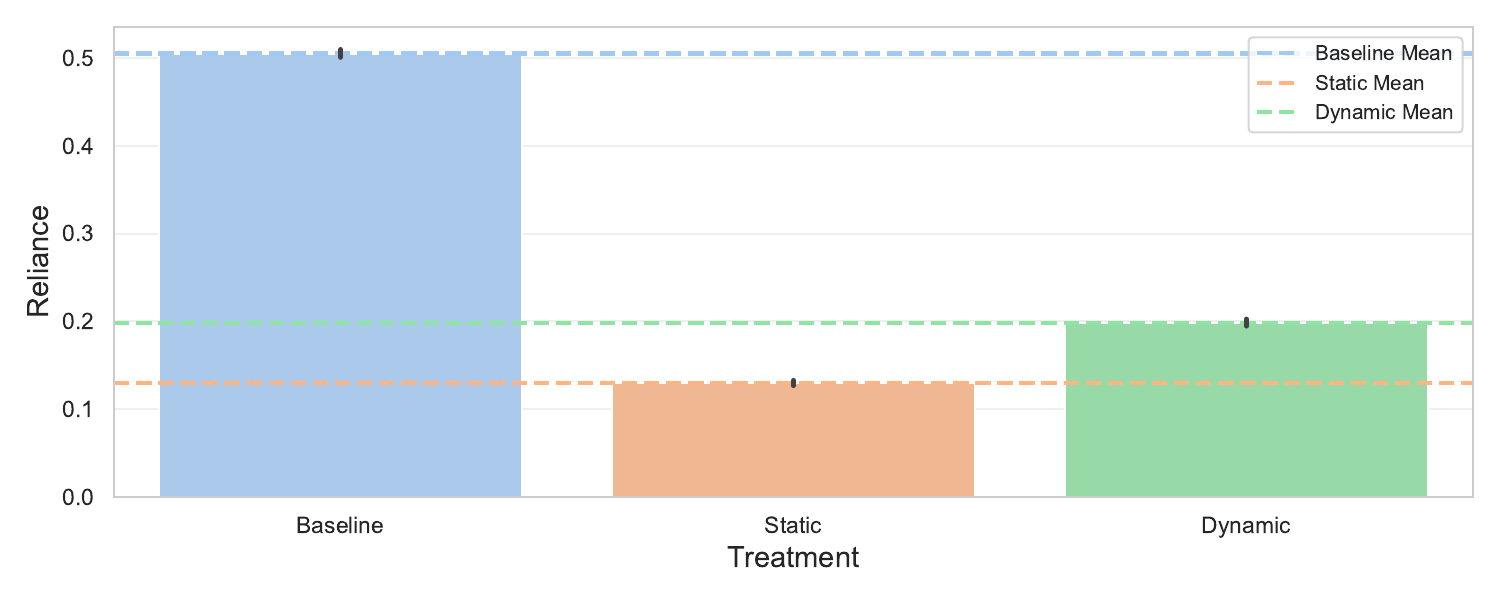}
 \caption{Weighted average participant reliance across treatments with 95\% confidence intervals.}
 \label{fig:reliance}
\end{figure}

Both treatments led to significant reductions in reliance compared to the baseline. Baseline 
participants exhibited relatively high reliance on AI advice (\(\mu_{b} = 0.505\)), while static 
bonus participants showed the lowest reliance (\(\mu_{s} = 0.130\); \(t = -133.48\), \(p < 
0.001\), \(d = -1.59\)), and dynamic bonus participants demonstrated intermediate levels of 
reliance (\(\mu_{d} = 0.199\); \(t = -97.06\), \(p < 0.001\), \(d = -1.34\)) (see 
\Cref{fig:reliance}). These results confirm that both incentive mechanisms reduced reliance on AI, encouraging more independent engagement with the task.

\textbf{Reliance Mitigation from Bonus Placement.} 
To understand how bonus placement shapes reliance behavior, we first examine whether the reduction in reliance within the dynamic treatment occurs specifically when bonuses are available. Participants showed significantly lower reliance on bonus-eligible instances (\(\mu_{bonus} = 
0.101\)) compared to non-bonus instances (\(\mu_{nobonus} = 0.452\); \(t = -76.93\), \(p < 
0.001\), \(d = -1.31\)). This indicates that reliance reductions are driven by bonus availability, reflecting targeted behavioral adaptation rather than a general treatment effect.

We next examine reliance behavior across instance configurations. \Cref{tab:reliance_instances} presents reliance rates for each scenario (S1–S3), showing that both incentive treatments reduced reliance across all instance types. Most importantly, both treatments significantly decreased reliance on instances where humans outperform AI (S3), effectively {encouraging human intervention when the AI is likely to be wrong}.

However, this shift came at a cost: both treatments also reduced reliance on instances where AI outperforms humans (S2), leading to \textit{underreliance} precisely when following AI advice would have been beneficial. Despite this, our earlier analysis showed that the dynamic bonus still improved overall team performance, suggesting that its targeted nature mitigates this trade-off more effectively than the static approach.

\begin{table}[h]
\centering
\resizebox{0.6\linewidth}{!}{%
\begin{threeparttable}
\begin{tabular}{lccc}
\toprule
\textbf{Instance Type} & \textbf{Baseline} & \textbf{Static} & \textbf{Dynamic} \\
\midrule
(S1) Human \& AI perform equally   & 0.485 & 0.177*** & 0.324*** \\
(S2) AI outperforms humans     & 0.734 & 0.297*** & 0.523*** \\
(S3) Humans outperform AI      & 0.221 & 0.154*** & 0.131*** \\
\bottomrule
\end{tabular}
\begin{tablenotes}
 \item \textit{Note:} * \textit{p} $<$ .05; ** \textit{p} $<$ .01; *** \textit{p} $<$ .001.
\end{tablenotes}
\end{threeparttable}
}
\vspace{0.5em}
\caption{Reliance rates across complementarity conditions.}
\vspace{-1em} 
\label{tab:reliance_instances}
\end{table}

\textbf{Gaming the System: Behavior Distortion.} Finally, we observe evidence of arbitrage behavior, where participants opted to ``solve" tasks independently but submitted answers that simply mirrored the AI's recommendation, an effortful detour taken solely to collect the bonus. This behavior was most pronounced in the static bonus treatment, where bonuses were awarded indiscriminately, regardless of AI confidence. Participants in this condition exhibited significantly higher rates of such behavior (\(\mu_s = 0.454\)) than those in the baseline (\(\mu_b = 0.173\); \(t = 42.80\), \(p < 0.001\), \(d = 1.68\)) or dynamic conditions (\(\mu_d = 0.318\); \(t = 19.43\), \(p < 0.001\), \(d = 0.75\)), where such gaming provided no advantage.

Qualitative responses further confirm this incentive-driven gaming. Participants explicitly described choosing to ``solve" tasks, even when they agreed with the AI, purely to receive the bonus: 
\textit{“Even if the AI's recommendation was correct, I would tend to solve the task myself in order to receive the bonus.”} 
\textit{“If I was pretty sure and the AI backed it up, then I would go ahead and take the additional bonus payment.”}

These findings reveal a critical downside of the static bonus: by overshooting and rewarding effort indiscriminately, it encourages arbitrary and unnecessary behavior, paradoxically increasing effort without improving outcomes. This underscores the delicate nature of incentive design: to be effective, bonuses must not only encourage engagement but must also be carefully targeted to avoid unintended consequences. The dynamic bonus, by tying rewards to the task-specific value of human input, better preserves this balance.

\section{Discussion}
\label{sec:discussion}
Despite the promise of human-AI collaboration, humans often overrely on AI advice, even when their own judgment would lead to better outcomes \citep{schemmer2022meta, vaccaro2024combinations}. In this work, we investigate how incentive structures shape this reliance on AI advice, revealing a fundamental misalignment in the economic incentives underlying human-AI collaboration. Our findings provide empirical validation for dynamic, task-aware incentives as a means to reduce overreliance and promote more effective collaboration.

\subsection{Key Findings}
Our primary theoretical contribution lies in identifying a previously overlooked structural cause for systematic overreliance by analyzing the framework established by \citet{bansal2021most}: the misalignment of incentives in AI-assisted decision-making. While existing research has focused extensively on cognitive factors such as trust and confidence calibration \citep{Kahr2024, ma2024you} as well as mental models \citep{bansal2019beyond, van_den_bosch_design_2025, holstein2025development}, our analysis reveals that the challenge may also be rooted in the economic incentive structures of human-AI collaboration design, positioning incentive misalignment as an additional, complementing driver of overreliance alongside existing cognitive factors \citep{bertrand2022cognitive, spitzer2025human}.

Our behavioral experiment subsequently validates the efficacy of the proposed incentive mechanisms, confirming that both static and dynamic bonuses reduce reliance on AI advice. However, their effects on human-AI team performance diverge. Static bonuses harmed performance by incentivizing independent effort indiscriminately, even when AI advice remains superior, and enabled strategic gaming, in which participants solve tasks independently to collect bonuses while mirroring AI advice. Dynamic bonuses, by contrast, improved performance by focusing incentives on instances where human judgment is most valuable, demonstrating that effective incentive design requires targeting the value of human input rather than independent effort alone.

\subsection{Implications}

\textbf{Addressing Overreliance Requires Targeting Structural Causes.} Our findings demonstrate that a key driver of overreliance is not inherent to either the human or the AI system, but rather stems from a misalignment in the incentive structures embedded in the system design. When humans are not appropriately incentivized to invest cognitive effort in verifying or overriding AI recommendations, they default to acceptance---even when human judgment would yield better outcomes. This suggests that effective human-AI collaboration cannot be achieved solely through improving AI accuracy or human training \citep{pinski2023ai, spitzer2025human}. Instead, it requires addressing the economic and structural conditions under which decisions are made. System designers should focus on creating incentive mechanisms that encourage humans to engage their independent judgment selectively and strategically.

\textbf{Incentives Must Be Context-dependent Rather Than Static.} The contrasting outcomes between static and dynamic bonuses highlight the importance of tailoring incentives to instances where human judgment is expected to add the most value. Static bonuses, applied uniformly, created opportunities for gaming, ultimately harming performance. In contrast, dynamic bonuses showed promise by reducing overreliance and improving overall decision quality. However, our findings also show that dynamic bonuses, while more targeted, can also lead to reduced reliance even when the AI is demonstrably superior. This suggests that although context-sensitive incentives are a step in the right direction, further refinements are needed. 

\textbf{Challenges of Designing Optimal Incentives.} Even within our simplified framework, designing an optimal bonus mechanism presents several challenges. Key variables such as human effort (\(\lambda\)) and accuracy (\(P_\text{H}\)) are difficult to observe, subjective, and vary across individuals and tasks. This complicates calibration and risks misaligned incentives. Static bonuses may oversimplify, while overly complex dynamic schemes may require reliable real-time estimates that are difficult to acquire. Obtaining well-calibrated AI confidence estimates is a practical prerequisite for the dynamic bonus mechanism, and while established techniques such as temperature scaling or isotonic regression exist for this purpose \citep{wang2023calibration}, their effectiveness varies across models and domains, requiring practitioners to verify calibration quality before deployment. Moreover, even well-intentioned bonuses can distort behavior, encouraging strategic gaming or underreliance in unintended situations. However, while we demonstrate that dynamic incentives improve human-AI team performance, we also find that these incentives increase cognitive load, resulting in lower task completion rates. This suggests that more sophisticated incentive mechanisms may have unintended consequences beyond their primary effects, requiring careful evaluation of trade-offs in real-world implementations. These findings underscore that an effective design must balance simplicity, fairness, and adaptability while aligning incentives with the true value of human input.

\textbf{Deployment and Governance Requirements.} Translating our findings into practice requires careful attention to implementation details and governance structures. Organizations must establish transparency mechanisms that inform workers when bonuses are available and how they are determined, enabling informed decisions about effort allocation. This includes transparent communication about the relationship between AI confidence levels and bonus eligibility in dynamic schemes as employed within our study. Additionally, organizations require strategies to identify unintended consequences such as gaming behaviors \citep{larkin2014cost, courty2004empirical}, excessive cognitive burden \citep{kalakoski2020effects}, or systematic disadvantaging of certain worker populations \citep{buyl2022tackling, poe2024conflict}. Worker input into incentive design decisions represents another critical requirement, as ethical deployment demands worker participation in determining acceptable tradeoffs between accuracy, cognitive load, and compensation.

\textbf{Domain Applicability and Incentive-Compatible System Design.} The effects of incentive misalignment vary across domains, and our theoretical model clarifies where overreliance is most problematic. Settings characterized by high cognitive effort costs ($\lambda$) or low incentive levels ($\gamma+\beta$) produce the largest misalignment term, meaning rational agents will defer to AI even when their own judgment would yield better outcomes. Medical diagnosis, legal review, and content moderation are  particularly concerning in this regard \citep{obermeyer2019dissecting, angwin2022machine},  as large caseloads and time pressure keep effort costs high, while performance incentives  in these settings may not adequately reflect the marginal value of independent judgment.  These domains also tend to have the audit infrastructure needed to implement performance-contingent bonuses, since individual decisions are logged and outcomes become observable over time, providing the basis for awarding $\theta$. By contrast, domains involving subjective quality judgments, real-time decisions under strict latency constraints, or outcomes observable only with long delays present greater implementation challenges, as do traditional employment settings with fixed salary structures. Identifying domains where both the problem and the remedy are most tractable therefore represents a promising direction for future work.

\subsection{Ethical Considerations}
Our findings raise important questions about the ethics of incentive design in AI-assisted work.

\textbf{Worker Welfare and Cognitive Burden.} While dynamic bonuses improved team performance, they significantly increased cognitive load and reduced task completion rates. This creates a tension that organizations must navigate thoughtfully. On the one hand, further increased cognitive engagement may pose challenges for workers, for example, with cognitive differences \citep{lysaght2008towards}, or non-native language speakers \citep{albl2020cognitive}. On the other hand, incentivizing active cognitive engagement may serve a protective function against skill degradation. Recent evidence suggests that passive reliance on AI systems, particularly generative AI, can lead to deskilling and erosion of expertise over time \citep{lee2025impact, Tankelevitch2025}. By encouraging workers to maintain cognitive involvement in their work, incentive mechanisms may support the long-term preservation and development of human capabilities \citep{ericsson2004deliberate}. The key question becomes not whether to impose cognitive demands, but how to design incentives that balance skill maintenance and task performance with worker capacity and well-being.

\textbf{Worker Agency and Distributive Fairness.} If dynamic bonuses increase organizational effectiveness while increasing workers' cognitive load, fair implementation requires ensuring that workers are adequately compensated for increased cognitive demands. In this context, incentive mechanisms may differentially affect worker populations. Moreover, bonuses tied to task difficulty require workers to accurately assess when their judgment adds value---a metacognitive skill that varies across individuals \citep{flavell1979metacognition}. For example, workers with less experience, AI literacy, or self-efficacy may struggle to consistently perform accurate self-assessments \citep{FERNANDES2026108779}.

\textbf{Performance-Centric Framing.} Our study focused on performance outcomes but did not assess other factors such as worker autonomy \citep{nie2015importance}, job satisfaction \citep{hemmer2023human}, or long-term well-being \citep{makikangas2016longitudinal}. This focus shaped our core framing, where we characterized excessive reliance on AI advice as problematic when human judgment would be more accurate. While accuracy maximization is a reasonable primary goal, the acceptable trade-off between accuracy and other goals varies across contexts. In lower-stakes contexts, workers may legitimately prefer some cognitive offloading even at modest accuracy costs. Conversely, in high-stakes domains like healthcare \citep{obermeyer2019dissecting} or criminal justice \citep{angwin2022machine}, accuracy and fairness must remain paramount regardless of worker preferences.

\subsection{Limitations and Directions for Future Work}

Our study shows that incentive structures systematically influence reliance behavior in human-AI collaboration. By isolating this mechanism in a controlled setting, we demonstrate a causal relationship between incentive design and the mitigation of overreliance. While this fundamental work was necessarily constrained by some factors that limit generalizability, these factors simultaneously open several promising directions for future research that can build upon and extend our findings.

\textbf{Extending Across Domains and Stakes.} We deliberately selected an image classification task that requires no specialized expertise, allowing us to isolate the effect of incentive mechanisms from domain-specific confounds. This methodological choice, common in behavioral research \citep{fugener2021will}, aims to ensure that observed effects stem from the incentive structure itself rather than task-specific confounds. Having established this baseline effect, future research should examine how incentive mechanisms operate across different domains and stakes. Professional decision-making contexts such as healthcare, criminal justice, or financial services involve specialized knowledge and serious consequences for errors \citep{obermeyer2019dissecting, angwin2022machine}. These factors may amplify or attenuate the effects we observe, and understanding these contextual moderators represents an important next step.

\textbf{Sample Characteristics.} While we aimed for balanced representation in terms of gender and age through our recruitment criteria, our Prolific sample may not represent workplace populations across other dimensions such as educational backgrounds, cognitive abilities, or professional experience. Workers in organizational settings face different motivational structures, such as career advancement \citep{bernadette2013understanding}, peer relationships \citep{grant2022social}, or intrinsic professional commitment \citep{gerhart2015pay, nie2015importance}, which our monetary incentives do not capture. Future research should investigate how these factors influence responses to incentive mechanisms in organizational contexts and explore how personalizing incentive mechanisms based on these factors affects reliance in human-AI collaboration.

\textbf{Longitudinal Effects and Adaptation.} Our short-term experimental session establishes immediate behavioral responses to incentive mechanisms. However, sustained exposure may produce adaptation effects, skill development or atrophy, or shifting motivations. Longitudinal studies examining whether initial behavioral changes persist, fade, or evolve over time would provide critical insights for practical implementation \citep{steyvers2024three}. Such work could reveal whether incentive mechanisms require periodic recalibration, whether workers develop more sophisticated strategies for effort allocation, or whether prolonged exposure leads to habituation effects that diminish treatment efficacy.

\textbf{Combining Incentive and Information-Based Interventions.} Our work demonstrates that incentive mechanisms can reduce overreliance by addressing structural misalignment in human-AI collaboration. However, prior research has shown that information-based interventions such as explanations \citep{buccinca2025contrastive}, confidence calibration \citep{ma2024you}, or specialized training \citep{pinski2023ai, spitzer2025human} can also improve reliance behavior. Future research should investigate whether combining incentive mechanisms with these approaches yields synergistic effects.

\textbf{Theoretical Framework Assumptions.} Our formal model assumes rational agents who maximize expected utility through a linear combination of rewards, penalties, bonuses, and effort costs. This simplification isolates incentive misalignment as a driver of overreliance, consistent with \citet{bansal2021most}, who themselves acknowledge that humans are not purely rational. For example, cognitive biases \citep{bertrand2022cognitive} or varying risk preferences \citep{yang2026my} may produce systematic deviations. Additionally, the relationships between $\gamma$, $\beta$, $\theta$, and $\lambda$ may interact in more complex, non-linear ways that our formulation does not capture, and future work should explore more complex utility formulations accounting for diminishing returns or interaction effects between effort costs and incentive magnitude.

\textbf{Boundary Conditions of Dynamic Incentives.} Our dynamic bonus mechanism relies on calibrated AI confidence as a practical proxy for the expected value of human effort, which remains observable without ground-truth human accuracy estimates. However, this proxy may be less reliable in settings where human and AI capabilities are unknown or rapidly shifting. Future work should examine how dynamic incentives perform under such conditions, including settings with multiple human decision-makers whose capabilities vary. Additionally, the confidence threshold distinguishing bonus-eligible from non-eligible instances was derived a priori from the theoretical framework, leaving room for future work to examine the sensitivity of the results to threshold variations across domains and AI systems.

\section{Conclusion}
\label{sec:conclusion}
We investigate how incentive structures influence overreliance in traditional human-AI collaboration setups and propose an alternative mechanism design to address systemic bias toward overreliance on AI advice. Our theoretical analysis uncovers misaligned incentive structures as a structural driver of overreliance. Our behavioral experiment with 180 participants then demonstrates that dynamic incentives simultaneously reduce overreliance and improve human-AI team performance. In contrast, static incentives foster strategic behavior that undermines their intended purpose. While our findings are demonstrated in image classification, the identified bias term may operate differently across domains with varying cognitive effort requirements and decision stakes. Future research should investigate how these mechanisms operate across diverse domains and populations, develop adaptive approaches that balance performance with worker welfare, and explore governance structures that ensure incentive designs serve both organizational goals and worker well-being.

\bibliographystyle{ACM-Reference-Format}
\bibliography{bibliography}

\newpage
\appendix

\section*{Appendix}
\label{sec:appendix}
This appendix provides detailed methodological specifications and supplementary analyses supporting our main findings. We present comprehensive implementation details for our behavioral experiment, including AI system training, instance selection criteria, and incentive mechanism derivations, followed by extended statistical analyses and robustness checks. These materials ensure experimental replicability in our analytical approach, particularly regarding our inverse variance weighting methodology for handling heterogeneous measurement precision across treatments.

\subsection*{Experimental Design}
\label{appendix:experimental_design}

\textbf{AI System Training Details.}\label{appendix:ai_system} We implemented a DenseNet-161 architecture pre-trained on ImageNet. The dataset was split into 60\% training, 10\% validation, and 20\% test sets. The model was fine-tuned on distorted images over 40 epochs using SGD optimizer with a learning rate of $1 \cdot 10^{-3}$, weight decay of $5 \cdot 10^{-4}$, and cosine annealing learning rate scheduler. We used a batch size of 32 and applied early stopping on the validation loss. The AI system achieved an accuracy of 67.5\% on the test set. 

As we later provide the binned confidence rating of the AI system to the humans, we aimed for a calibrated confidence design. Specifically, we binned the confidence in four levels ranging from very low (0-0.25), over low (0.25-0.5) and high (0.5-0.75) to very high (0.75-1).

\textbf{Instance Selection.}
Effective evaluation of human-AI collaboration requires understanding when each partner should contribute to decision-making. In real-world scenarios, the relative capabilities of humans and AI systems vary across different task instances, creating distinct strategic considerations for optimal collaboration. When both partners perform similarly on a task instance, efficiency considerations favor following AI advice due to reduced cognitive effort costs. When AI demonstrates superior performance, optimal collaboration requires humans to rely on AI advice. However, at instances where humans outperform AI, humans should invest effort to solve task instances independently to achieve better outcomes rather than (over)relying on AI advice.

Given these theoretical foundations, we strategically selected instances to represent these three complementary scenarios: (S1) instances where both humans and AI perform similarly, (S2) instances where AI outperforms humans, and (S3) instances where humans outperform AI. Our goal was to construct a balanced dataset enabling the evaluation of how different incentive mechanisms perform across the full spectrum of realistic scenarios of human-AI collaboration.

To operationalize these theoretical categories, we leveraged two complementary indicators: human annotator agreement patterns from \citet{steyvers2022bayesian} and calibrated confidence ratings:

\textbf{Annotator Agreement.} We reasoned that human annotator disagreement serves as a proxy for human probability for solving a task correctly with high disagreement indicating instances that humans find challenging, while low disagreement suggests instances where humans can reliably perform well.

\textbf{Calibrated Confidence.} We used temperature scaling to calibrate the confidence of the AI. In detail, we optimized the temperature parameter \(T\) on the validation set and divided the logits by the selected parameter prior to the softmax to infer the predictions on the test set. This allowed us to target accuracy rates of AI predictions that correspond to the midpoint of each confidence bin: ``very low'' confidence (0-0.25 confidence range, 12.5\% target accuracy), ``low'' confidence (0.25-0.5 confidence range, 37.5\% target accuracy), ``high'' confidence (0.5-0.75 confidence range, 62.5\% target accuracy), and ``very high'' confidence (0.75-1.0 confidence range, 87.5\% target accuracy). 

\begin{figure}
  \centering
  \includegraphics[width=0.6\linewidth]{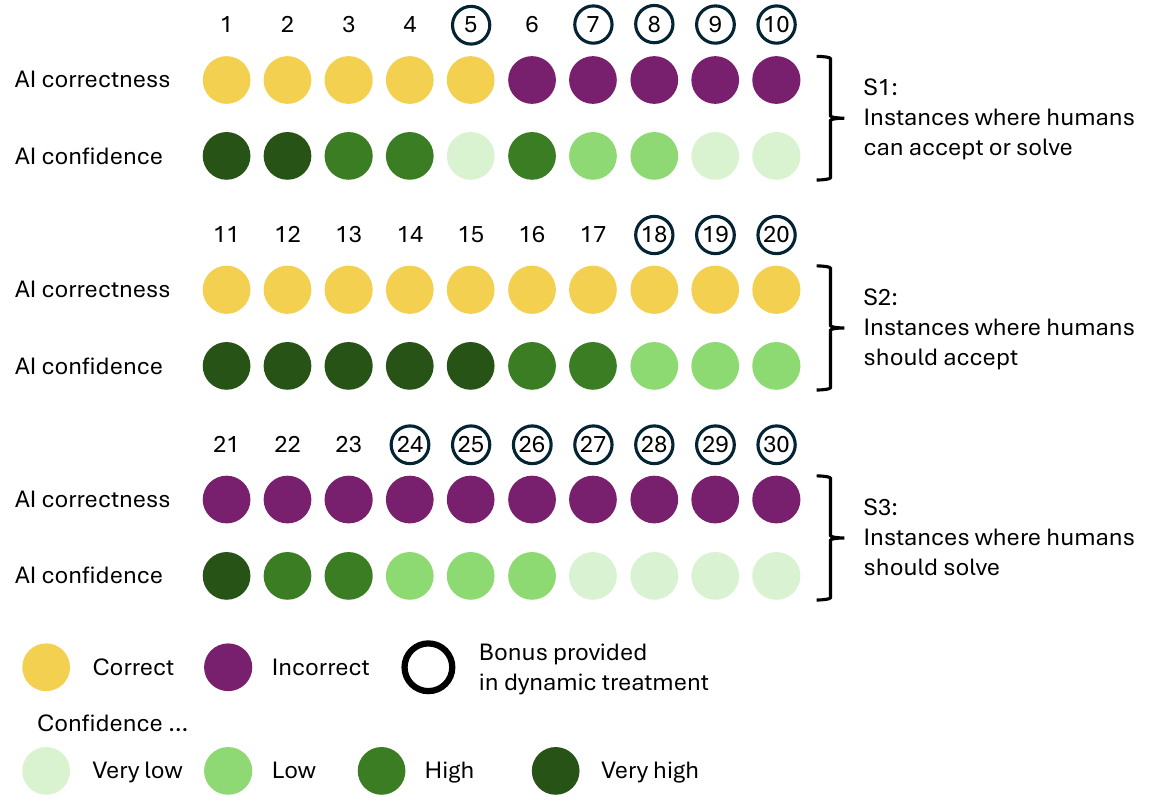}
  \caption{Distribution of instances with confidences to complementary subsets.}
  \label{fig:instances_confidence_calibration}
\end{figure}

Building on this logic, we categorized instances by selecting instances with high human annotator disagreement (a minimum of 4 out of 6 annotators disagreeing) compared to instances with strong human consensus (a maximum of 3 out of 6 annotators disagreeing). Additionally, instances can be classified based on whether the AI was correct/incorrect in its predictions. Strategically sampling from these options, allows us to select instances for S1-S3, ensuring that our instance selection aligns with the underlying capability differences that define each complementarity scenario. \Cref{fig:instances_confidence_calibration} demonstrates our instance selection, including how confidence levels distribute across our three complementarity subsets, confirming the coherence of our selection methodology.

\begin{figure}[h]
 \centering
 \includegraphics[height=0.35\textheight]{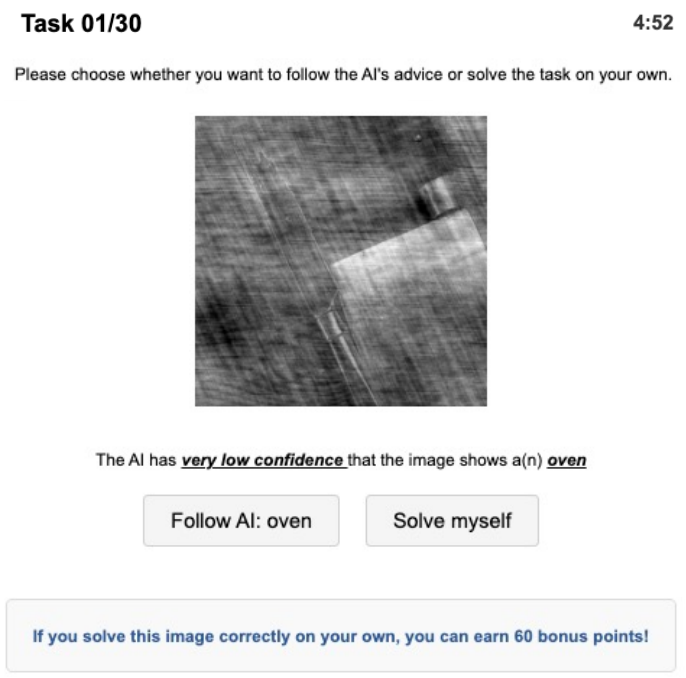}
 \includegraphics[height=0.35\textheight]{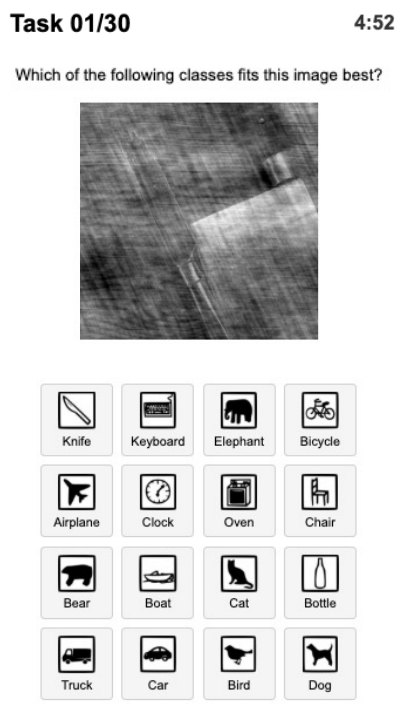}
 \caption{Task interface showing the main decision screen (left) and the independent classification screen (right). The bonus notification (shown in blue) was absent in the baseline condition. In the static condition, a fixed bonus of 30 points was displayed for all instances; in the dynamic condition, the bonus varied by instance---60 points for low AI confidence instances and no bonus for high AI confidence instances---directing human effort to where it was most valuable, with every 10 points translating to \text{\text{£}}0.01.}
 \label{fig:task_interface}
\end{figure}

\textbf{Incentive Mechanism Design.}
\label{appendix:incentive_design}
In the following, we outline how we determined the incentive mechanisms across treatments. The overall mechanism was designed to maintain equivalent maximum payouts across all treatments while systematically varying the distribution between rewards ($\gamma$) and independent decision-making bonuses ($\theta$).

The maximum variable performance-based payment for each participant can be expressed as:
\begin{align*}
\text{Var Payment}_{\max} = \sum_{i=1}^{n} (\gamma + \theta_i \cdot \mathbb{I}[\text{solve}_i])
\end{align*}
where $\mathbb{I}[\text{solve}_i]$ is an indicator function equal to 1 if the participant chooses to solve instance $i$ independently and answers correctly, and 0 otherwise.

Based on pilot testing, we anticipated a study duration of approximately 10 minutes, leading to a base participation payment of \pounds1.00 to ensure fair compensation above minimum wage standards. Additionally, our theoretical framework necessitates performance-based incentives to motivate human effort in decision-making. To create meaningful incentives for task engagement while maintaining experimental feasibility, we set the maximum performance-based payment at \pounds1.80, resulting in a maximum total compensation of \pounds2.80 for an estimated task duration of 10 minutes, which corresponds to a high remuneration according to Prolific standards. Following, we outline the theoretical derivation of the distribution of the performance-based payment across ($\gamma$) and ($\theta$).

\textbf{Baseline.} In traditional AI-assisted decision-making settings, participants receive rewards based solely on final decision accuracy, regardless of whether they rely on AI advice or solve independently. This represents the standard incentive structure that our theoretical analysis identifies as systematically biased toward overreliance. With $\theta = 0$, rearranging the maximum variable payout results in:
\begin{align*}
\gamma_{\text{b}} &= \frac{\text{Var Payment}_{\max}}{n_{\text{instances}}} = \frac{\text{£}1.80}{30} = \text{£}0.06
\end{align*}

\textbf{Static.} For the static bonus setting, we construct two comparable scenarios to align the incentives for two extreme behavioral strategies: (E1) solving all instances independently and (E2) always following AI advice. Our goal is to ensure that a rational decision-maker would be indifferent between these two strategies in expectation. To make the comparison meaningful, we incorporate time constraints reflective of our experimental setup. Let \( t_{\text{max}} \) denote the total time allowed for the task (300 seconds in our experiment), and \( t_{\text{per instance}} \) the average time required for a human to solve one instance independently (20 seconds, based on \citet{hemmer_complementarity_2025}). Additionally, let \( p_{\text{H}} \) denote the average human accuracy and \( p_{\text{AI}} \) the average AI accuracy, both computed against ground-truth labels from empirical data from \citet{steyvers2022bayesian}.

For humans always rejecting AI advice and solving all instances independently (E1), we receive:
\begin{align*}
\text{Performance}_{\text{E1}} &= \frac{t_{\text{max}}}{t_{\text{per instance}}} \cdot p_{\text{H}}
\end{align*}

For humans always following AI advice without independent solving (E2):
\begin{align*}
\text{Performance}_{\text{E2}} &= n_{\text{instances}} \cdot p_{\text{AI}}
\end{align*}

Setting equal expected payouts between strategies:

\begin{align*}
(\gamma_{s} + \theta_{s}) \cdot \text{Performance}_{\text{E1}} &= \gamma_{s} \cdot \text{Performance}_{\text{E2}} \\
\theta_{s} &= \gamma_{s} \cdot \left(\frac{\text{Performance}_{\text{E2}}}{\text{Performance}_{\text{E1}}} - 1\right)
\end{align*}

Substituting our experimental parameters ($t_{\text{max}} = 300$ seconds, $t_{\text{per instance}} = 20$ seconds, and $p_{\text{H}} = p_{\text{AI}} = 0.5$ based on our balanced instance selection) results in:

\begin{itemize}
  \item Performance$_{\text{E1}} = 300s / 20s \cdot 0.5 = 7.5$ instances
  \item Performance$_{\text{E2}} = 30 \cdot 0.5 = 15$ instances
\end{itemize}

This yields the equivalence of $\theta_{\text{s}} = \gamma_{\text{s}}$. Given the maximum payment constraint, we receive:
\begin{align*}
\text{£}1.80 &= 30 \cdot (\gamma_{\text{s}} + \theta_{\text{s}}) \\
\gamma_{\text{s}} = \theta_{\text{s}} &= \text{£}0.03
\end{align*}

\textbf{Dynamic.} The dynamic condition strategically allocates bonuses to instances where human judgment provides the highest expected utility. Unlike the static bonus setting with a single bonus level, we implement a dual-bonus structure with two different bonus amounts ($\theta_{\text{d, 1}}$ and $\theta_{\text{d, 2}}$) applied conditionally based on AI confidence levels, while maintaining overall payment equivalence across treatments.

The performance-based payment structure becomes:
\begin{align*}
\text{Var Payment}_{\max} = \sum_{i=1}^{n} (\gamma_{d} + &\theta_{d, 1,i} \cdot \mathbb{I}[\text{solve}_i \wedge p_{\text{AI},i} \geq 0.5] + \\ &\theta_{d, 2,i} \cdot \mathbb{I}[\text{solve}_i \wedge p_{\text{AI},i} < 0.5])
\end{align*}

For the dynamic treatment, we preserve equivalent maximum variable payment relative to the static condition ($\sum_{i=1}^{30} (\gamma + \theta_i) = \text{£}1.80$), with instances distributed such that half satisfy $p_{\text{AI}} < 0.5$ and half satisfy $p_{\text{AI}} \geq 0.5$. To focus the entire bonus budget on instances where human input is most valuable, we set $\theta_{\text{d, 1}} = \text{£}0.00$ for high-confidence instances ($p_{\text{AI}} \geq 0.5$) and maintain $\gamma_{\text{d}} = \text{£}0.03$ (same as static condition).

\begin{align*}
\text{£}1.80 &= 30 \cdot \gamma_{\text{d}} + 15 \cdot \theta_{\text{d, 1}} + 15 \cdot \theta_{\text{d, 2}}
\end{align*}

Substituting $\gamma_{\text{d}} = \text{£}0.03$ and $\theta_{\text{d, 1}} = \text{£}0.00$:
\begin{align*}
\theta_{\text{d, 2}} &= \text{£}0.06
\end{align*}

This yields:
\begin{align*}
\gamma_{\text{d}} &= \text{£}0.03 \\
\theta_{\text{d, 2}} &= \text{£}0.06 \quad \text{(low AI confidence: } p_{\text{AI}} < 0.5\text{)} \\
\theta_{\text{d, 1}} &= \text{£}0.00 \quad \text{(high AI confidence: } p_{\text{AI}} \geq 0.5\text{)}
\end{align*}

This effectively doubles the incentive for independent effort when human judgment adds the most value.

This design creates an hourly compensation range of \pounds11.40-\pounds13.80 based on individual performance and task completion rates, ensuring fair compensation while maintaining strong incentives for strategic decision-making across all experimental conditions. This range is derived from our empirical dataset analysis and instance selection design. Based on the collected labels \citep{steyvers2022bayesian}, we expect humans to correctly solve approximately 50\% of instances independently, establishing the lower bound of our compensation structure. Under the different payment structures, this baseline human performance translates to an average minimum payment of \pounds11.40 per hour.
The upper bound (\pounds13.80) reflects human-AI complementarity, where participants can potentially solve 25 out of 30 instances correctly. This scenario accounts for the 5 instances in our dataset where neither humans nor AI systems typically perform well.

\newpage

\subsection*{Experimental Results}
In the following, we elaborate our testing methodology in more detail.

\subsubsection*{Testing Methodology}
\label{appendix:analysis}

Due to identified systematic differences in task completion rates across treatments, participants provided varying numbers of observations, creating heterogeneous measurement precision across our two primary outcome measures: human-AI team performance, and reliance behavior. Participants completing fewer instances yield less reliable estimates due to smaller sample sizes, violating the equal-variance assumption underlying standard analytical approaches.

Following \citet{hartung2011statistical}, we employed \textit{inverse variance weighting }to address this analytical challenge, i.e., assigning greater weight to observations with higher precision (lower variance) and reduced weight to less reliable measurements (higher variance). For each participant $i$, we calculated weights based on the inverse of their estimated measurement variance:

$$w_i = \frac{1}{\sigma_i^2}$$

The specific variance calculation depends on the outcome measure, i.e., human-AI team performance and reliance. As both are binary outcomes, both measures, when aggregated, follow a binomial distribution. While instance difficulties vary, randomized presentation ensures that each participant encounters a representative sample of instances, allowing us to model their overall performance with success parameter \(p_i\) as a binomial distribution. This results in:

$$w_i = \frac{n_i}{p_i(1-p_i)}$$

where $n_i$ represents the number of instances classified by participant $i$ and $p_i$ is the proportion of instances where the participant followed AI advice. For performance measures, we calculated sample variance based on individual accuracy across completed instances.

This weighting scheme creates computational difficulties when participants exhibit perfect reliance behavior (either $p_i = 0$ or $p_i = 1$), as the denominator $p_i(1-p_i)$ approaches zero, resulting in infinite weights. To address this issue while preserving the high information value, we assigned participants with either $p_i = 0$ or $p_i = 1$ a weight equal to 125\% of the maximum finite weight observed among other participants within the same treatment group. Results were robust to alternative scaling factors (100\%-200\% of maximum finite weight). This approach ensures these participants receive appropriately high but finite influence in the analysis.

To improve robustness and mitigate the influence of extreme weights, we applied winsorization \citep{tukey1977exploratory} at the 5th and 95th percentiles to the weight distributions within each treatment group before conducting comparative analyses. This procedure caps extreme weights while preserving the overall weighting structure, preventing any single participant from disproportionately influencing results.

\textit{Statistical Testing.} All comparative analyses for human-AI team performance and reliance rates employed two-sided Welch's t-tests with inverse variance weighting to account for heterogeneous task completion rates across participants unless otherwise specified. Given our sample size of 60 participants per treatment condition, we proceeded with t-tests without formal normality testing, leveraging the robustness of t-tests to departures from normality, particularly with moderate to large sample sizes \citep{sawilowsky1992more}.

\textit{Multiple Comparisons Correction.} To control for Type I error inflation arising from multiple treatment comparisons against the baseline condition, we applied Bonferroni correction to all p-values. Specifically, for each outcome measure, we conducted two primary comparisons (Static vs. Baseline and Dynamic vs. Baseline), resulting in an adjusted p-values for individual tests.

\subsubsection*{Task Completion}
\label{appx: task_completion}
Task completion rates differed across treatments, with static bonus participants completing the most instances ($\mu_s = 26.933$), followed by baseline ($\mu_b = 25.350$), and dynamic bonus participants ($\mu_d = 23.633$). Shapiro-Wilk tests indicated non-normal distributions across all conditions ($p < 0.001$), while Levene's test revealed heterogeneous variances ($p = 0.012$).

Kruskal-Wallis tests confirmed significant differences in completion rates across conditions ($H = 7.505$, $p = 0.023$). One-sided Kolmogorov-Smirnov tests revealed no significant distributional shift for static vs. baseline ($D = 0.117$, $p = 0.407$), a trend for dynamic vs. baseline ($D = 0.200$, $p = 0.091$), and a significant shift between static vs. dynamic conditions ($D = 0.233$, $p = 0.038$), with static participants completing approximately 14\% more instances than dynamic participants.

\textit{Mediation Analysis.} We conducted mediation analysis to examine whether cognitive load mediates the relationship between treatments and task completion, using bootstrap analysis with 5000 simulations (\Cref{tab:mediation_full}). The analysis demonstrates a significant Average Causal Mediation Effect (ACME = -0.441, 95\% CI [-1.129, 0.000], p = 0.048), with cognitive load accounting for approximately 26\% of the dynamic treatment effect, indicating that dynamic bonuses operate substantially through increased cognitive burden.

\subsubsection*{Human-AI Team Performance}
Performance analyses employed the same inverse variance weighting methodology for both overall treatment comparisons across the full dataset (\Cref{tab:performance_overall_stats}) and subset analyses across three complementarity scenarios (S1-S3, \Cref{tab:performance_detailed_stats}).

\textit{Influence of Bonus Availability on human-AI team performance.} To examine the targeted effects of bonus placement, we conducted conditional analyses comparing dynamic treatment participants against baseline participants on identical instance subsets (\Cref{tab:performance_bonus_efficacy}). Specifically, we partitioned instances based on bonus availability in the dynamic condition and compared performance between treatments within each partition. This within-instance comparison controls for instance-specific difficulty effects and isolates the impact of bonus availability.

\subsubsection*{Reliance Behavior}
Reliance analyses employed the same inverse variance weighting methodology for overall treatment comparisons across the full dataset (\Cref{tab:reliance_overall_stats}) and subset analyses across complementarity scenarios (S1-S3, \Cref{tab:reliance_detailed_stats}). 

\textit{Influence of Bonus on Reliance Behavior.} To examine whether bonus availability directly influences reliance behavior, we conducted within-subjects analyses comparing dynamic treatment participants' reliance on bonus-eligible versus non-bonus instances (\Cref{tab:reliance_bonus_efficacy}). This paired design controls for individual differences while isolating the effect of bonus availability. For participants with different completion patterns across bonus and non-bonus instances, weights were combined using the harmonic mean.

\textit{Arbitrage Behavior Analysis.} We also examined instances where participants chose to solve independently but submitted answers matching AI recommendations---a form of strategic gaming to collect bonuses without genuine independent effort. Specifically, we compared static treatment against both baseline and dynamic treatment followed by appropriate correction of p-values (\Cref{tab:arbitrage_stats}).

\begin{table*}
\centering
\renewcommand{\arraystretch}{1.3}
\begin{tabular}{l r r r r} 
\hline
Dependent variable & \multicolumn{2}{c}{Cognitive Load} & \multicolumn{2}{c}{Task Completion} \\
\cmidrule(lr){2-3} \cmidrule(lr){4-5}
& coeff & se & coeff & se \\
\hline \hline
Intercept & 4.178*** & 0.128 & 29.920*** & 2.157 \\
\multicolumn{5}{l}{Treatment:} \\
\quad \textit{- Baseline} & & & & \\
\quad \textit{- Static} & 0.019 & 0.181 & 1.605 & 1.152 \\
\quad \textit{- Dynamic} & 0.403* & 0.181 & -1.276 & 1.168 \\
Cognitive Load & & & -1.094* & 0.478 \\
\hline
$R^2$ & \multicolumn{2}{c}{0.034} & \multicolumn{2}{c}{0.071} \\
$F$-statistic & \multicolumn{2}{c}{3.146*} & \multicolumn{2}{c}{4.483**} \\
\hline
\end{tabular}

\vspace{0.6em}

\begin{tabular}{l r r r r} 
\hline
& \multicolumn{4}{c}{Bootstrap Mediation Effects (5000 simulations)} \\
\cmidrule{2-5}
& Effect & Boot SE & Boot LLCI & Boot ULCI \\
\hline \hline
\multicolumn{5}{l}{Dynamic versus Baseline:} \\
\quad ACME & -0.441* & 0.288 & -1.129 & 0.000 \\
\quad ADE & -1.276 & 1.255 & -3.702 & 1.210 \\
\quad Total Effect & -1.717 & 1.247 & -4.114 & 0.760 \\
\quad Prop. Mediated & 0.257 & 0.983 & -1.541 & 2.250 \\
\hline
\end{tabular}
\begin{tablenotes}
\centering
\item[1] \textit{Note:} *\textit{p} $<$ .05; ** \textit{p} $<$ .01; *** \textit{p} $<$ .001.
\end{tablenotes}
\caption{Mediation analysis: Effect of incentive structures on efficiency through cognitive load.}
\label{tab:mediation_full}
\end{table*}

\begin{table*}
\centering
\begin{threeparttable}
\begin{tabular}{lcccccc}
\toprule
\textbf{Comparison} & \textbf{Treatment} & \textbf{Baseline} & \textbf{t} & \textbf{p} & \textbf{d} \\
& \textbf{Mean} & \textbf{Mean} & & & \\
\midrule
Static vs. Baseline & 0.632 & 0.643 & -6.008 & $<$0.001 & -0.101 \\
Dynamic vs. Baseline & 0.653 & 0.643 & 5.042 & $<$0.001 & 0.087 \\
\bottomrule
\end{tabular}
\end{threeparttable}
\caption{Overall treatment effects on human-AI team performance.}
\label{tab:performance_overall_stats}
\end{table*}

\begin{table*}[h]
\centering
\begin{threeparttable}
\begin{tabular}{lcccccc}
\toprule
\textbf{Condition} & \textbf{Dynamic} & \textbf{Baseline} & \textbf{t} & \textbf{p} & \textbf{d} \\
& \textbf{Mean} & \textbf{Mean} & & & \\
\midrule
Bonus Available & 0.537 & 0.524 & 2.832 & 0.009 & 0.069 \\
No Bonus Available & 0.796 & 0.786 & 1.517 & 0.259 & 0.031 \\
\bottomrule
\end{tabular}
\end{threeparttable}
\caption{Effect of bonus availability on human-AI team performance.}
\label{tab:performance_bonus_efficacy}
\end{table*}

\begin{table*}
\centering
\begin{threeparttable}
\begin{tabular}{lccccccccc}
\toprule
\textbf{Instance Type} & \textbf{Baseline} & \textbf{Static} & \textbf{Dynamic} & \multicolumn{3}{c}{\textbf{Static vs. Baseline}} & \multicolumn{3}{c}{\textbf{Dynamic vs. Baseline}} \\
\cmidrule(lr){5-7} \cmidrule(lr){8-10}
& \textbf{Mean} & \textbf{Mean} & \textbf{Mean} & \textbf{t} & \textbf{p} & \textbf{d} & \textbf{t} & \textbf{p} & \textbf{d} \\
\midrule
(S1) Human \& AI perform equally & 0.613 & 0.601 & 0.599 & -2.371 & 0.036 & -0.065 & -2.330 & 0.040 & -0.066 \\
(S2) AI outperforms humans & 0.814 & 0.841 & 0.871 & 6.137 & $<$0.001 & 0.127 & 12.788 & $<$0.001 & 0.264 \\
(S3) Humans outperform AI & 0.499 & 0.482 & 0.573 & -2.305 & 0.042 & -0.059 & 9.692 & $<$0.001 & 0.251 \\
\bottomrule
\end{tabular}
\end{threeparttable}
\caption{Human-AI team performance across complementarity conditions.}
\label{tab:performance_detailed_stats}
\end{table*}

\begin{table*}
\centering
\begin{threeparttable}
\begin{tabular}{lcccccc}
\toprule
\textbf{Comparison} & \textbf{Treatment} & \textbf{Baseline} & \textbf{t} & \textbf{p} & \textbf{d} \\
& \textbf{Mean} & \textbf{Mean} & & & \\
\midrule
Static vs. Baseline & 0.130 & 0.505 & -133.48 & $<$0.001 & -1.586 \\
Dynamic vs. Baseline & 0.199 & 0.505 & -97.06 & $<$0.001 & -1.336 \\
\bottomrule
\end{tabular}
\end{threeparttable}
\caption{Overall treatment effects on reliance behavior.}
\label{tab:reliance_overall_stats}
\end{table*}

\begin{table*}
\centering
\begin{threeparttable}
\begin{tabular}{lcccccc}
\toprule
\textbf{Condition} & \textbf{With Bonus} & \textbf{Without Bonus} & \textbf{t} & \textbf{p} & \textbf{d} \\
& \textbf{Mean} & \textbf{Mean} & & & \\
\midrule
Dynamic Treatment & 0.101 & 0.452 & -76.93 & $<$0.001 & -1.309 \\
\bottomrule
\end{tabular}
\begin{tablenotes}
 \item \textit{Note:} Comparison within dynamic treatment participants.
\end{tablenotes}
\end{threeparttable}
\caption{Effect of bonus availability on reliance behavior.}
\label{tab:reliance_bonus_efficacy}
\end{table*}

\begin{table*}
\centering
\begin{threeparttable}
\begin{tabular}{lccccccccc}
\toprule
\textbf{Instance Type} & \textbf{Baseline} & \textbf{Static} & \textbf{Dynamic} & \multicolumn{3}{c}{\textbf{Static vs. Baseline}} & \multicolumn{3}{c}{\textbf{Dynamic vs. Baseline}} \\
\cmidrule(lr){5-7} \cmidrule(lr){8-10}
& \textbf{Mean} & \textbf{Mean} & \textbf{Mean} & \textbf{t} & \textbf{p} & \textbf{d} & \textbf{t} & \textbf{p} & \textbf{d} \\
\midrule
(S1) Human \& AI perform equally & 0.485 & 0.177 & 0.324 & -47.31 & $<$0.001 & -1.160 & -21.04 & $<$0.001 & -0.519 \\
(S2) AI outperforms humans & 0.734 & 0.297 & 0.523 & -63.59 & $<$0.001 & -1.356 & -27.50 & $<$0.001 & -0.635 \\
(S3) Humans outperform AI & 0.221 & 0.154 & 0.131 & -13.14 & $<$0.001 & -0.273 & -20.32 & $<$0.001 & -0.447 \\
\bottomrule
\end{tabular}
\end{threeparttable}
\caption{Reliance behavior across complementarity conditions.}
\label{tab:reliance_detailed_stats}
\end{table*}

\begin{table*}
\centering
\begin{threeparttable}
\begin{tabular}{lcccccc}
\toprule
\textbf{Comparison} & \textbf{Static} & \textbf{Comparison Group} & \textbf{t} & \textbf{p} & \textbf{d} \\
& \textbf{Mean} & \textbf{Mean} & & & \\
\midrule
Static vs. Baseline & 0.454 & 0.173 & 42.80 & $<$0.001 & 1.682 \\
Static vs. Dynamic & 0.454 & 0.318 & 19.43 & $<$0.001 & 0.749 \\
\bottomrule
\end{tabular}
\end{threeparttable}
\caption{Strategic gaming (arbitrage) behavior: Static treatment comparisons.}
\label{tab:arbitrage_stats}
\end{table*}

\clearpage

\section*{Acknowledgements}
Generative AI tools were utilized throughout this work. Specifically, Claude and Github Copilot were employed to generate code for visualizations. Additionally, ChatGPT, DeepL Write, and Grammarly were used to enhance the writing quality of tutorials and explanations provided to participants during the experiments, as well as to improve the language throughout this paper.

\end{document}